\documentclass[journal]{IEEEtran}

\usepackage{graphicx,eucal}

\ifCLASSINFOpdf
\else
\fi
\begin{document}
\title{Quantum Sensor Miniaturization}

\author{G. Gilbert, M. Hamrick, Y. S. Weinstein, S. P. Pappas and A. Donadio
%\author{Michael~Shell,~\IEEEmembership{Member,~IEEE,}
%        John~Doe,~\IEEEmembership{Fellow,~OSA,}
%        and~Jane~Doe,~\IEEEmembership{Life~Fellow,~IEEE}% <-this % stops a space
%\thanks{M. Shell is with the Department
%of Electrical and Computer Engineering, Georgia Institute of Technology, Atlanta,
\thanks{The authors are with the {\sc MITRE} Quantum Information Science Group, MITRE, 260 Industrial Way West, 
Eatontown, NJ, 07724, USA e-mail: \{ggilbert, mhamrick, weinstein, spappas, tdonadio \}@mitre.org}}% <-this % stops a space
\maketitle

\begin{abstract}
%\boldmath
The classical bound on image resolution defined by the Rayleigh limit can be beaten by exploiting the properties of quantum mechanical entanglement. If entangled photons are used as signal states, the best possible resolution is instead given by the Heisenberg limit, an improvement proportional to the number of entangled photons in the signal. In this paper we present a novel application of entanglement by showing that the resolution obtained by an imaging system utilizing separable photons can be achieved by an imaging system making use of entangled photons, but with the advantage of a smaller aperture, thus resulting in a smaller and lighter system. This can be especially valuable in satellite imaging where weight and size play a vital role. 
\end{abstract}


% \fi
\IEEEpeerreviewmaketitle

\section{Introduction}
% I wish you the best of success.

%\hfill January 11, 2007

%\subsection{Subsection Heading Here}
%Subsection text here.
%\subsubsection{Subsubsection Heading Here}
%Subsubsection text here.
%\IEEEpubidadjcol

\IEEEPARstart{Q}{uantum} mechanical systems admit certain correlations that are 
not classically defined. Physical systems that exhibit these non-classical correlations 
are said to be entangled. Entanglement plays a central role in many emerging quantum 
technologies such as quantum computing \cite{NC}, quantum communications and cryptography 
\cite{Ekert}, and quantum interferometry \cite{Lloyd}. Applications of quantum 
interferometry include quantum lithography \cite{Boto}, quantum geodesy, 
quantum microscopy, {\it etc.} In this letter we introduce a novel application of 
photonic quantum interferometry, in which the properties of entanglement can be exploited 
to achieve a reduction in the size and weight of optical imaging systems. 

Previously proposed applications of photonic quantum interferometry rely on 
an effective reduction in diffraction that is exhibited by certain entangled states. 
This effective diffraction reduction can be exploited to achieve greater resolution in 
lithography \cite{Boto} and imaging \cite{Dowling}. The best possible imaging 
resolution that can be obtained using non-entangled photons is given by the Rayleigh 
limit 
\begin{equation}
R_R \simeq \frac{\lambda}{D},
\label{rr}
\end{equation}
where $\lambda$ is the wavelength of the light used by the imaging system and $D$ is 
the diameter of the aperture. 
Imaging systems that employ entangled photons can beat this resolution limit and 
achieve the so-called Heisenberg limit
\begin{equation}
R_H \simeq \frac{\lambda}{DN}
\label{rh}
\end{equation}
where $N$ is the number of entangled photons in the signal. 

In this paper we stipulate a scenario in which the resolution obtainable with unentangled light, 
constrained by the Rayleigh limit, is sufficient for the presumed purpose at hand. In this case we can exploit entanglement by trading resolution improvement 
for a corresponding reduction in size and weight of imaging optics.
Specifically, the use of signal states comprised of $N$ entangled photons allows 
for a reduction of the diameter of the imaging aperture by a factor of $N$ and a 
corresponding reduction in the volume and weight.
A reduction in the size and weight of the imaging system 
may be especially valuable in satellite imaging where weight and size 
play a vital role. 

\section{Photonic $N00N$ States and Quantum Sensor Miniaturization}

A particular entangled state that can achieve the above defined Heisenberg 
limit is the so-called $N00N$ state. A $N00N$ state is an $N$-photon path-entangled 
state. Given two spatial paths ($A$ and $B$) the $N00N$ state is given by 
\begin{equation}
|\psi_{N00N}\rangle = \frac{1}{\sqrt 2} \left( |N_A0_B\rangle  + |0_AN_B\rangle \right).
\end{equation}
Upon measurement, all $N$ photons will be observed in the same path, either $A$ or $B$,
with equal probability.

\begin{figure}[t]
\centering
\includegraphics[height=1.5in]{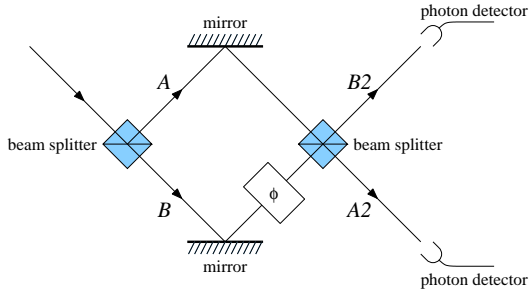}
\caption{Mach-Zehnder interferometer. A phase shift of $\phi$ in path $B$ can result from the presence of an object in the path, or from unequal path lengths between $A$ and $B$, or both. For unentangled light the probability of photon detection in the two photon detectors is $P \propto 1\pm\cos\phi$. For $N$ entangled $N00N$ state photons the probability of photon detection in the two photon detectors is $P \propto 1\pm\cos N\phi$.}
\label{mz}
\end{figure}

As noted above, the standard application of photonic quantum interferometry is to 
improve the image resolution by a factor of $N$.
We can illustrate the increased resolution that can be achieved with $N00N$ states by 
considering photon interference in a Mach-Zehnder interferometer. Figure \ref{mz} 
depicts a Mach-Zehnder interferometer in which a relative phase difference 
$\phi$ accumulates in path $B$. The relative phase difference can arise as the result 
of the presence of an object in the path, or of a difference in path length, or both. 
In the case of a single photon sent through the interferometer the detection 
probabilities of the two paths are given, respectively, by $P_{A2} \propto 1+\cos\phi$ and 
$P_{B2} \propto 1-\cos\phi$. For $N00N$ states, however, the amount of accumulated 
phase shift is proportional to $N$ such that the state of the system is given by
\begin{equation}
|\psi\rangle = \frac{1}{\sqrt 2} \left( |N_A0_B\rangle  + e^{iN\phi}|0_AN_B\rangle \right).
\end{equation}
The corresponding detection probabilities are thus given by 
$P_{A2} \propto 1+\cos N\phi$ and $P_{B2} \propto 1-\cos N\phi$. 
In this way we see that $N00N$ states exhibit an effective wavelength 
$\lambda_{N00N} = \lambda/N$, which results in the above claimed 
$N$-fold enhancement in resolution \cite{JBCY}. We point out that, in order to take advantage of the effective wavelength associated to $N00N$ states, it is neccesary to utilize a suitable $N$-entangled photon detection technique. This can be accomplished, for example, by making use of coincidence detection and appropriate post-processing of signal data, or by making use of suitable sensors that exhibit signatures that are directly responsive to $N$-photon processes.

We now demonstrate that resolution enhancement can be traded off for a
reduction in imaging system size. Although we utilize $N$-photon
entangled signal states, we nevertheless choose to remain constrained by the
Rayleigh limit on resolution,
\begin{equation}
R_R(D) \simeq \frac{\lambda}{D},
\label{rrd}
\end{equation}
where we now emphasize the dependence on aperture size, $D$.  As noted
above the use of entangled photonic signal states achieves the
Heisenberg limit on resolution given by
\begin{equation}
R_H(D) \simeq \frac{\lambda}{N\,D}.
\label{rhd}
\end{equation}
Even though we utilize entangled signal states, our objective is to
retain the resolution, $R_R$, that arises in classical imaging
systems.  We see that this can be accomplished by replacing the
aperture of size $D$ with a smaller aperture of size $D' = D/N$:
\begin{eqnarray}
R_H(D') & \simeq & \frac{\lambda}{N\,(D/N)} \nonumber \\
        & \simeq & R_R(D)
\label{rhdn}
\end{eqnarray}
Thus we have achieved our objective of obtaining image resolution
equivalent to the classical Rayleigh limit but with the advantage of a
smaller aperture.  This allows us to correspondingly reduce the size
and weight of our imaging system.
We will refer to this exploitation of entanglement to achieve a
reduction in the size of an imaging system as {\em quantum sensor
miniaturization}.

Let us now roughly estimate possible size and weight savings that can
be achieved with quantum sensor miniaturization.  We consider a simple 
model of a telescope as our test case.  
Our model telescope consists of an optic
at the end of a light weight skeleton support cage.  
We assume a cylindrical
geometry with diameter $D$ chosen as needed to
resolve features of interest.  

As shown above, the use of entangled $N$-photon signals enables us to
reduce the aperture diameter from $D$ to $D/N$, thus reducing the cross-sectional
area of the optic by a factor of $N^2$.  This reduces the mass of
the optic by the same factor, neglecting for simplicity the reduced
thickness needed to ensure that the optic maintain its figure against 
its own weight. For entangled photon signals with $N = 2$,
this leads to a reduction in weight by at least a factor of $4$ due 
to the reduced size of the optic.

As a specific example we now consider the LIDAR In-space Technology Experiment 
(LITE) which was flown on the Discovery shuttle in 1994 \cite{LITE}. 
The LITE system employed a 1-meter diameter telescope as a 
receiver. The LITE system transmitter was 
a Nd:YAG laser the output of which was doubled and tripled to produce 
wavelengths of 1064 nm, 532 nm, and 355 nm. 
Let us consider a modified version of the LITE system in which
entangled photonic states with $N = 2$ are used as signals.
We will refer to the LITE system modified in this way as `entangled-LITE.' LIDAR is a remote sensing technology, analogous to radar, that makes use of laser signals, rather than radio waves, to perform ranging and other measurements on irradiated targets. The use of optical wavelengths provides advantages over the use of radar in many circumstances. The use of $N00N$-state signals in entangled-LITE, rather than the standard laser signals used in ordinary LITE, retains the advantages of the use of LIDAR over radar. Entangled light, in this application, functions in much the same way as separable states of light, with relatively few modifications needed to enable a meaningful system improvement. In this scenario, the argument leading to eq.(7) indicates that the fiducial system resolution
can be maintained while halving the size of the telescope optic, leading to 
an approximate weight reduction by a factor of 4. 
In addition to considerations involving the optic, 
one can explore various possible laser sources in connection with an
entangled-LITE system including Nd:YAG, Ti:Sapphire, or any of the variety of 
violet (Ga-N) to cyan lasers currently available. Regarding the source of entangled photons, different spontaneous parametric downconversion 
crystals might be considered as well, including lithium 
triborate, barium borate, or bismuth borate. Additional equipment 
for entangled-LITE would include the hardware needed
to confirm the detection of entangled photons, such as $N$-photon detectors, coincidence circuits, {\it etc}. A notional comparison of 
these parameters is given in Table \ref{LITE}.

% An example of a floating table. Note that, for IEEE style tables, the 
% \caption command should come BEFORE the table. Table text will default to
% \footnotesize as IEEE normally uses this smaller font for tables.
% The \label must come after \caption as always.
%
\begin{table}[!t]
\renewcommand{\arraystretch}{1.3}
% if using array.sty, it might be a good idea to tweak the value of
% \extrarowheight as needed to properly center the text within the cells
\caption{LITE versus entangled-LITE}
\label{LITE}
\centering
% Some packages, such as MDW tools, offer better commands for making tables
% than the plain LaTeX2e tabular which is used here.
\begin{tabular}{|c||c|c|}
\hline
 & LITE & entangled-LITE \\
\hline\hline
T/R Optics & 1 meter telescope & .5 meter telescope \\\hline
Lasers & Nd:YAG & Nd:YAG, Ti:Sapph, etc. \\\hline
Crystal & C$^*$DA, KD$^*$P & BBO, BiBO, LBO \\\hline
Imaging Hardware & non-entangled light & $N = 2$ $N00N$-states \\\hline
\end{tabular}
\end{table}

It has recently been shown that the entanglement of photonic $N00N$ states 
is very susceptible to degradation by atmospheric attenuation.
Photonic $N00N$ states lose their entanglement and hence cannot achieve
the Heisenberg limit in the presence of atmospheric attenuation
and, moreover, this effect becomes worse as $N$ increases \cite{GHW}. This 
is an especially severe problem if the goal is to achieve enhanced image 
resolution, since one needs large values of $N$ to achieve meaningful 
improvement. (Moreover, the relatively modest increase of resolution 
obtainable for small $N$ is unlikely to be worth the cost of 
replacing current systems.) However, the atmospheric degradation of 
$N00N$ states is a less severe problem if the goal is to achieve quantum 
sensor miniaturization. This is because it is possible to obtain meaningful 
reductions in size and weight even for values of $N$ as small as $N=2$. Thus, 
quantum sensor miniaturization can be viable in the presence of atmospheric 
attenuation, even in circumstances for which $N00N$-state-based entanglement-enhanced 
image resolution is not viable. For completeness, we note that there are alternative types of entangled states besides path-entangled states, and it may be useful for specific realizations of imaging sensor hardware to make use of entangled photon states other than photonic $N00N$ states. Finally, we point out that both $N00N$-state-based quantum sensor miniaturiztion, as well as $N00N$-state-based entanglement-enhanced image resolution should offer significant improvements in exo-atmospheric scenarios in which atmospheric attenuation does not play a role.

\section{Conclusion}

In this letter we have shown that the special features inherent in the entanglement correlations 
of photonic $N00N$ states can be exploited to achieve reductions in the size and weight of optics 
employed in imaging systems: we refer to this process as quantum sensor miniaturization. We showed 
that meaningful improvements in size and weight can be achieved for imaging through the atmosphere 
upon utilization of quantum sensor miniaturization, with values of $N$ as small as $N=2$. This is 
the case in spite of the fact that atmospheric degradation limits the viability of photonic $N00N$ 
states for the purpose of achieving improvement in image resolution, where large values of $N$ are 
required. We noted that both image resolution ehnhancement, and quantum sensor miniaturization, 
may be quite viable in exo-atmospheric scenarios. Quantum sensor miniaturization may prove extremely 
valuable for satellite imaging where weight and size are at a premium.

\section*{Acknowledgment}

This research was supported under MITRE Technology Program Grant 20MSR912.
The authors would like to thank A. Bram for useful comments and input.

\ifCLASSOPTIONcaptionsoff
  \newpage
\fi



%\bibliographystyle{IEEEtran}

\end{document}